\documentclass[epjCONF]{svjour}
\usepackage{graphics}
\usepackage{graphicx}
\usepackage[varg]{txfonts} 
\usepackage[latin1]{inputenc}
\usepackage{hyperref} 
\usepackage[pdftex,dvipsnames,usenames]{color}		

\DeclareGraphicsExtensions{.pdf, .jpg, .eps}

\def\beq{\begin{equation}}
\def\eeq{\end{equation}}
\def\bea{\begin{eqnarray}}
\def\eea{\end{eqnarray}}

\def\eqref#1{Eq.~(\ref{eq:#1})}

\newcommand*{\figref}[1]{Fig.~\ref{fig:#1}}
\newcommand*{\figlab}[1]{\label{fig:#1}}

\def\VYP#1#2#3{{\bf #1} (#2) #3}  
\def\ApP#1#2#3{Astropart.~Phys.~\VYP{#1}{#2}{#3}}
\def\NIM#1#2#3{Nucl.~Inst.~Meth.~\VYP{A#1}{#2}{#3}}

\def\PRL#1#2#3{Phys.~Rev.~Lett.~\VYP{#1}{#2}{#3}}

\newcommand{\etal}{\mbox{\textit{et al.}}}                       %

\session-title{UHECR2012}
\begin{document}
\title{What the radio signal tells about the cosmic-ray air shower.}
\author{Olaf Scholten\inst{1}\fnmsep\thanks{\email{scholten@kvi.nl}} \and Krijn D. de Vries\inst{1} \and Klaus Werner\inst{2} }
\institute{KVI, University of Groningen, Groningen, The Netherlands \and SUBATECH, Nantes, France}
\abstract{
The physics of radio emission from cosmic-ray induced air showers is shortly summarized.  It will be shown that the radio signal at different distances from the shower axis provides complementary information on the longitudinal shower evolution, in particular the early part, and on the distribution of the electrons  in the shower core.  This complements the information obtained from surface, fluorescence, and muon detectors and is very useful in getting a comprehensive picture of an air shower.
} 
\maketitle
\section{Introduction}
\label{intro}

There are several approaches that are followed to model radio emission from extensive air showers. These can crudely be separated into two categories, Microscopic and Macroscopic. In a Microscopic approach the tracks of the individual electrons are followed and the emitted radiation of each is summed to yield the total radiation of the extensive air shower. This approach is incorporated in the REAS~\cite{REAS} and the ZHAireS~\cite{ZHaireS} codes. In the alternative, Macroscopic, approach the velocity distribution of the electrons is summed locally to yield the macroscopic charge and current density distribution in the shower. From this four-current distributions the (relativistic) Maxwell equations~\cite{Jac-CE} are used to generate the radiated fields. This approach is followed in the MGMR~\cite{MGMR,Wer08} and EVA~\cite{EVA} calculations. Up to a few years ago the two approaches showed large inconsistencies which were shown to be mainly due to a rather subtle term that was missing in the Microscopic calculations. By correcting this, consistency could be reached~\cite{Hue11} which is a major achievement and indicates that, based on first principles, we basically understand the emission process. It also shows that a dual approach is necessary to be able to trust the results. It should be noted that for coherent radiation the two approaches should give the same results, they differ in the incoherent contribution to the radiation field.

The main advantage of a macroscopic approach is that it clearly indicates the aspects of the shower dynamics that are responsible for certain features of the detected radio signal. The macroscopic approach requires a parametrization of the position and time dependence of the four-currents in the shower which can be performed at various levels of sophistication. In the following we use the Macroscopic approach. Since radio emission is basically governed by wave mechanics any features in the frequency spectrum of the pulse are directly related to critical length scales in the distribution of the particles in the shower.

\section{Basics of Radio-wave emission}
\label{sec:1}

To understand radio emission from extensive air showers it is important to keep in mind that a constant current (at a fixed distance) does not radiate. To create radiation one needs a varying current and/or a distance to this current that changes. In an air shower we have both, the charge/current distribution is concentrated in the shower which rushes towards the surface of the Earth with the velocity of light in vacuum (denoted by $c$). In addition, the current and charge densities are (roughly) proportional to the particle density in the shower which grows from the first collision to reach a maximum at $X_{max}$ whereafter it decreases rapidly again. Since the latter introduces a fast time dependence in the currents this is the dominant driving mechanism for the emission of MHz radiation.

An equally important consideration is the fact that coherent radiation can only be emitted from current and charge densities, involving many particles, that have a typical spatial and temporal extent that is smaller than  the wave-length or the inverse-frequency of the radiation.
With this simple principle the main features of the pulse shape can be understood~\cite{Sch10}. Here we will go a step further and focus on the influence of the finite refractivity of air, ${\cal N}=n-1$, on the measured radio signal as this turns out to be essential to come to a correct interpretation.

\section{The Macroscopic model}

The particles in the cosmic-ray shower move through the magnetic field of Earth with the light velocity $c$. The lighter ones, electrons and positrons, will receive a substantial acceleration due to the action of the Lorentz force. This acceleration is counteracted by the collisions with the ambient air molecules and results in an (air-pressure dependent) drift velocity with opposite directions for electrons and positrons, $v_d\approx 0.04\,c$. As a result a net electric current develops which is proportional to the number of particles in the shower and oriented in the direction of the Lorentz force, $\hat{x}=-\vec{v} \times \vec{B}$, where $\vec{v}$ is the direction of the original cosmic ray. This varying current is responsible for -so called- geomagnetic radiation which is polarized along the direction of the emitting current, $\hat{x}$.

In addition to the electric current there is also a net charge excess in the shower due to the knock-out of electrons from air molecules by elastic positron-electron collisions and Compton scattering. In shower simulations the charge excess is approximately equal to 20\% of the number of particles and is thus substantial, and, equally important, varies with shower height $z$ (the distance, measured along the shower, to the point of impact on Earth). Because of this variation there is also charge-excess radiation which is radially polarized and as such distinguishable from geomagnetic radiation~\cite{Rev11,Schoorl11,Berg12}.

\section{Index of refraction}

\begin{figure}[!htb]
\centerline{ \includegraphics[width=0.45\columnwidth, keepaspectratio]{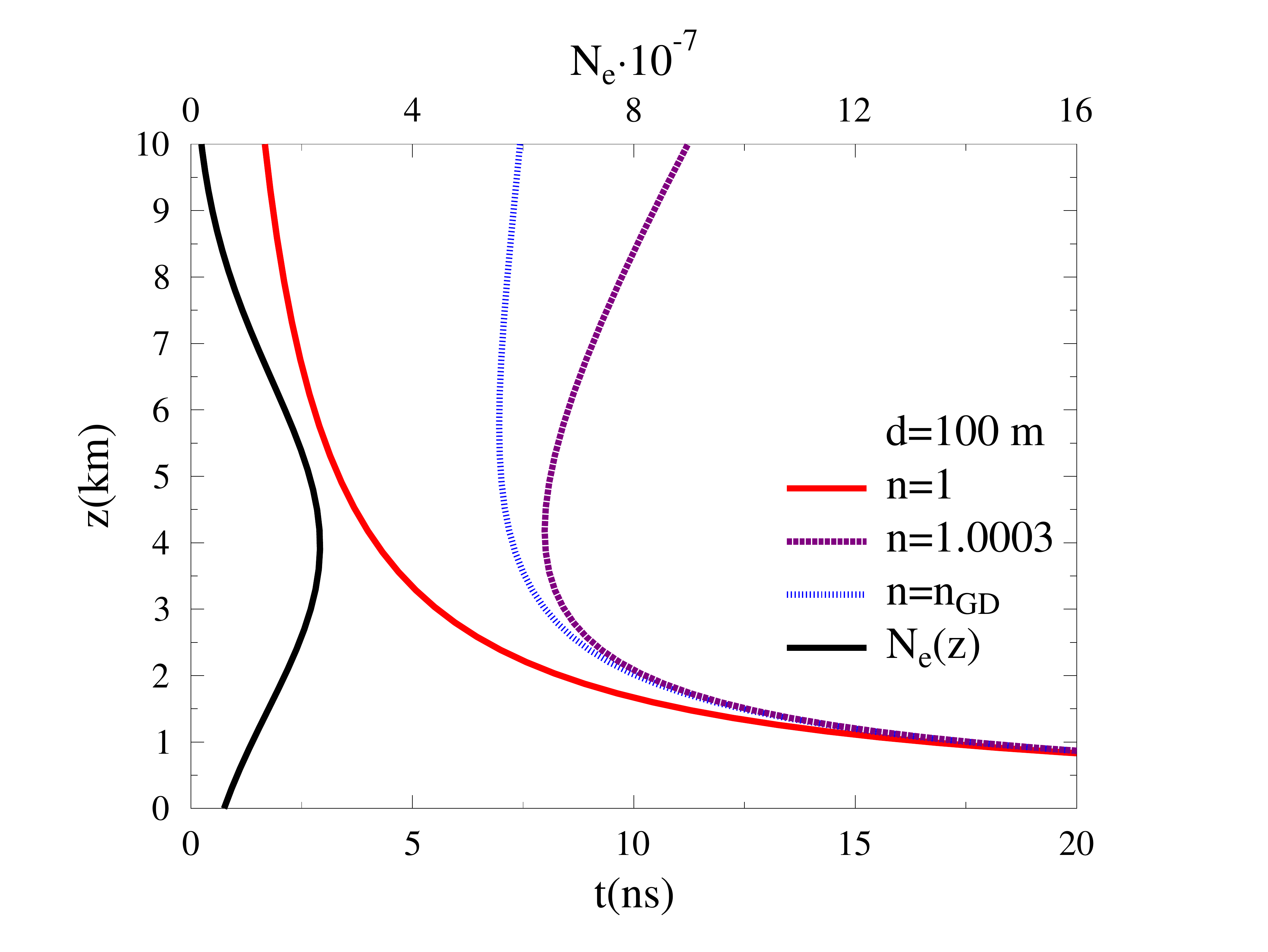} \includegraphics[width=0.48\columnwidth, keepaspectratio]{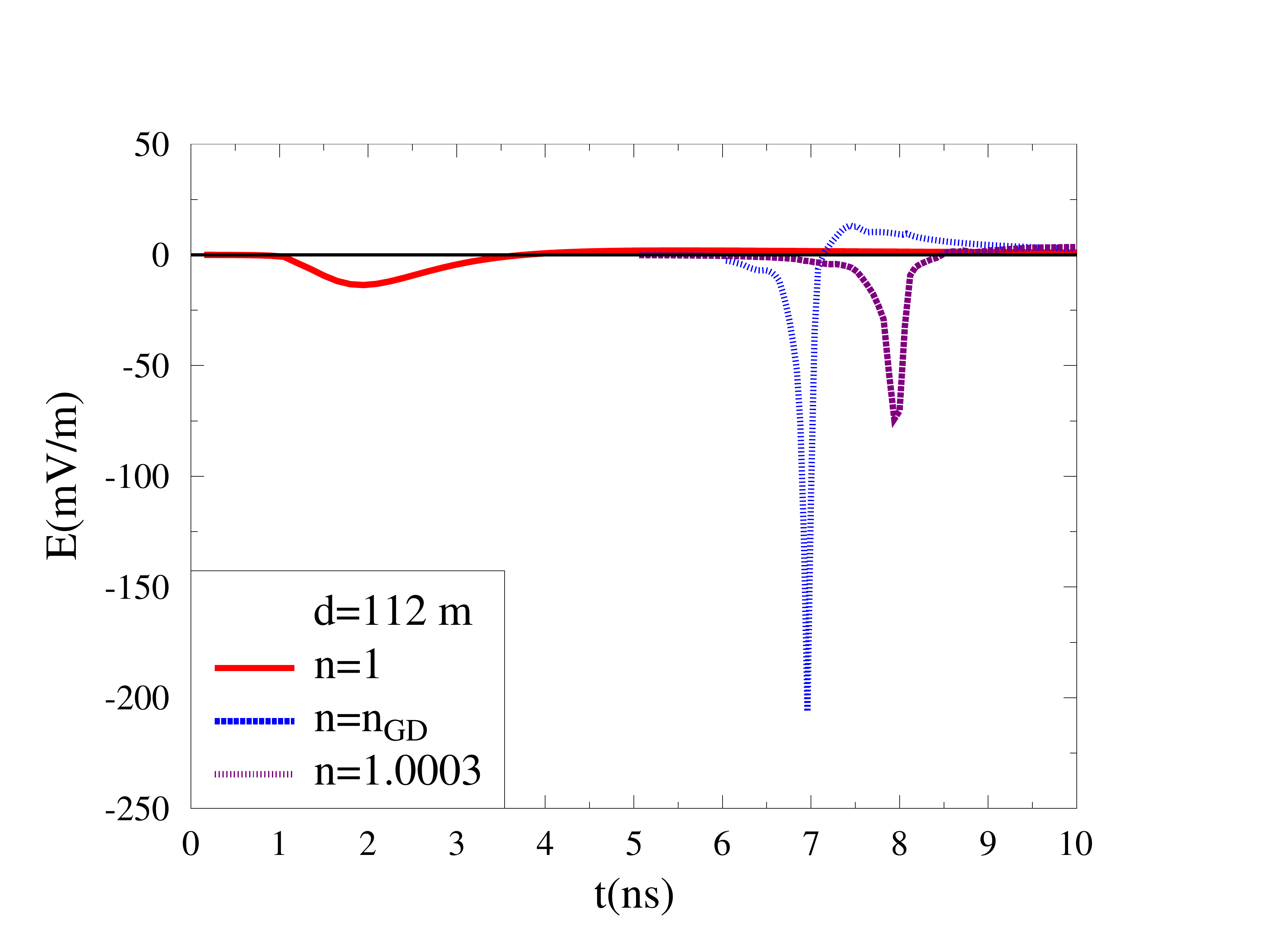}}
\caption{L.h.s.: the emission height $z$ as function of the observer time $t$ for three different values of the index of refraction for an observer at a distance of 112\,m from the shower core (impact distance is 100\,m). The dashed line gives the shower-profile as function of $z$ for a $E=5\times 10^{17}$\,eV proton-induced shower at $\theta=27^\circ$. The calculated pulses for the three cases of the index of refraction $n$ (vacuum, constant Earth-surface value, realistic air-density dependence) are shown on the r.h.s.}
\figlab{trett}
\end{figure}

Of particular importance to understand the effect of the finite refractivity of air is the relation between the time, $t$, for an observer at a distance $d$ from the shower core and the height, $z$, from which this part of the signal was emitted. For the simple geometry of a vertical shower it can be shown that $t \approx d^2/2cz$. This implies that signals emitted from the top part of the shower arrive earlier than those emitted from later stages. In \figref{trett} (l.h.s.) the red-colored curve shows the dependence between emission height and observer time for the case of vanishing refractivity. It can easily be seen that this follows closely the approximate formula $t \approx d^2/2cz$. The black curve drawn in \figref{trett} gives the vertical shower profile. As can be deduced from this picture the leading edge of the resulting pulse lies around $t=1$\,ns with a zero crossing of the pulse (related to the shower maximum) near $t=4$\,ns, consistent with the pulse shown on the r.h.s.\ of \figref{trett}.

Including a realistic refractivity of air ($n=n_{GD}$ indicating a dependence on air density as given by the law of Gladstone and Dale) changes the picture in an essential way as shown by the light blue dotted curves in \figref{trett}. Due to the fact that the shower propagates faster than radio waves, much of the radiation emitted along a finite part of the shower arrives at the observer at the same time. This is clear from the l.h.s.\ of the figure by the fact that the blue curve maps all points with $z>4$\,km to the same observer time of $t=7$\,ns. As a result the pulse (on the r.h.s.) shows a very strong peak at this time. The finite but very narrow width of the peak (as compared to the one for $n=1$) is due to the fact that the important-limiting length scale is no longer the projected length of the shower (which tends to zero at this point) but the distribution of the particles in the shower front. Simulations show that close to the shower core the particles in the longitudinal direction are concentrated within length-scales of mm's while in the transverse direction this may be meters.
Projecting this transverse distance along the line of sight, because of the angles involved, very short dimensions emerge translating in a very narrow pulse.

\begin{figure}[!htb]
\centerline{ \includegraphics[width=0.45\columnwidth, keepaspectratio]{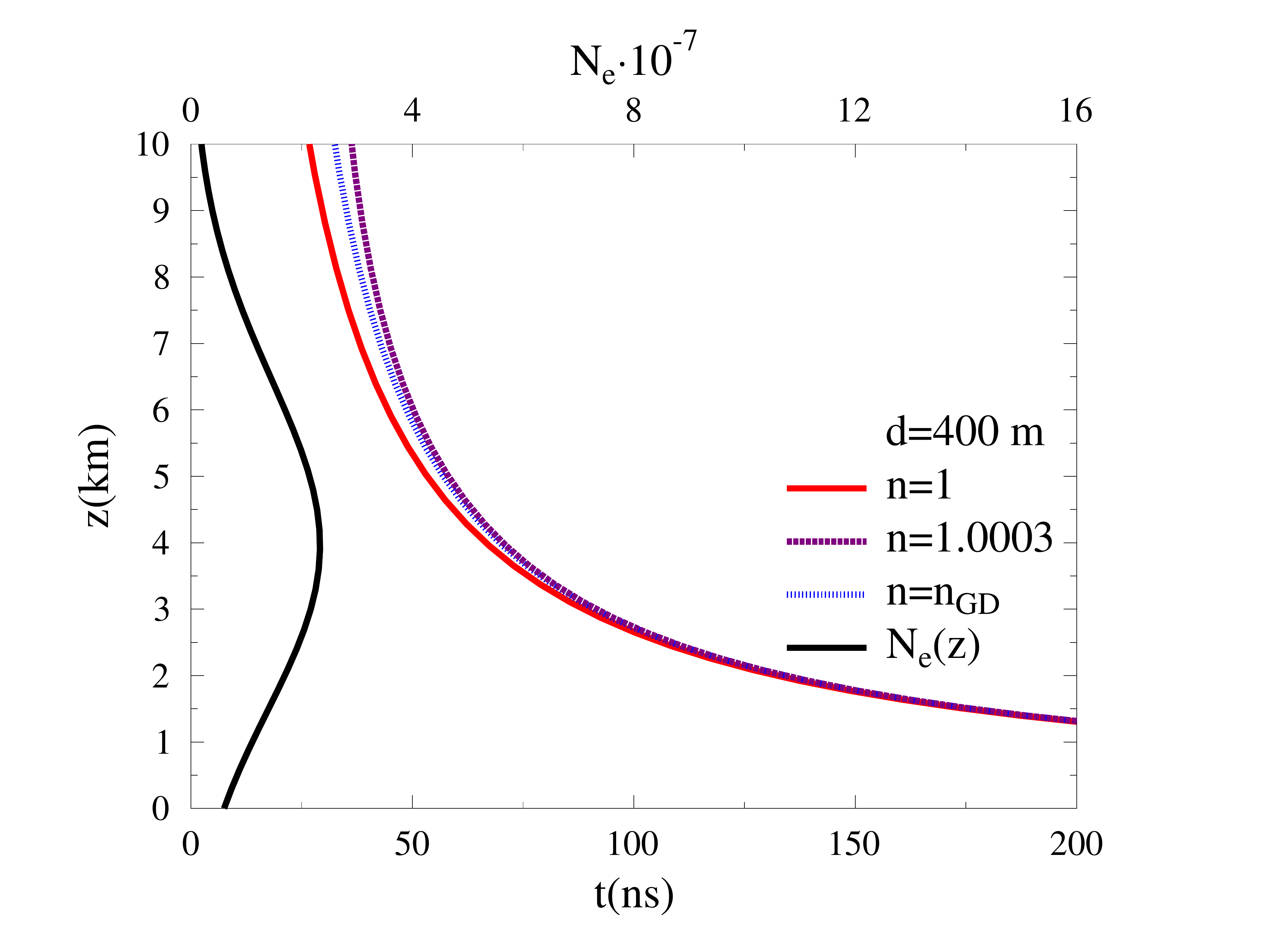} \includegraphics[width=0.5\columnwidth, keepaspectratio]{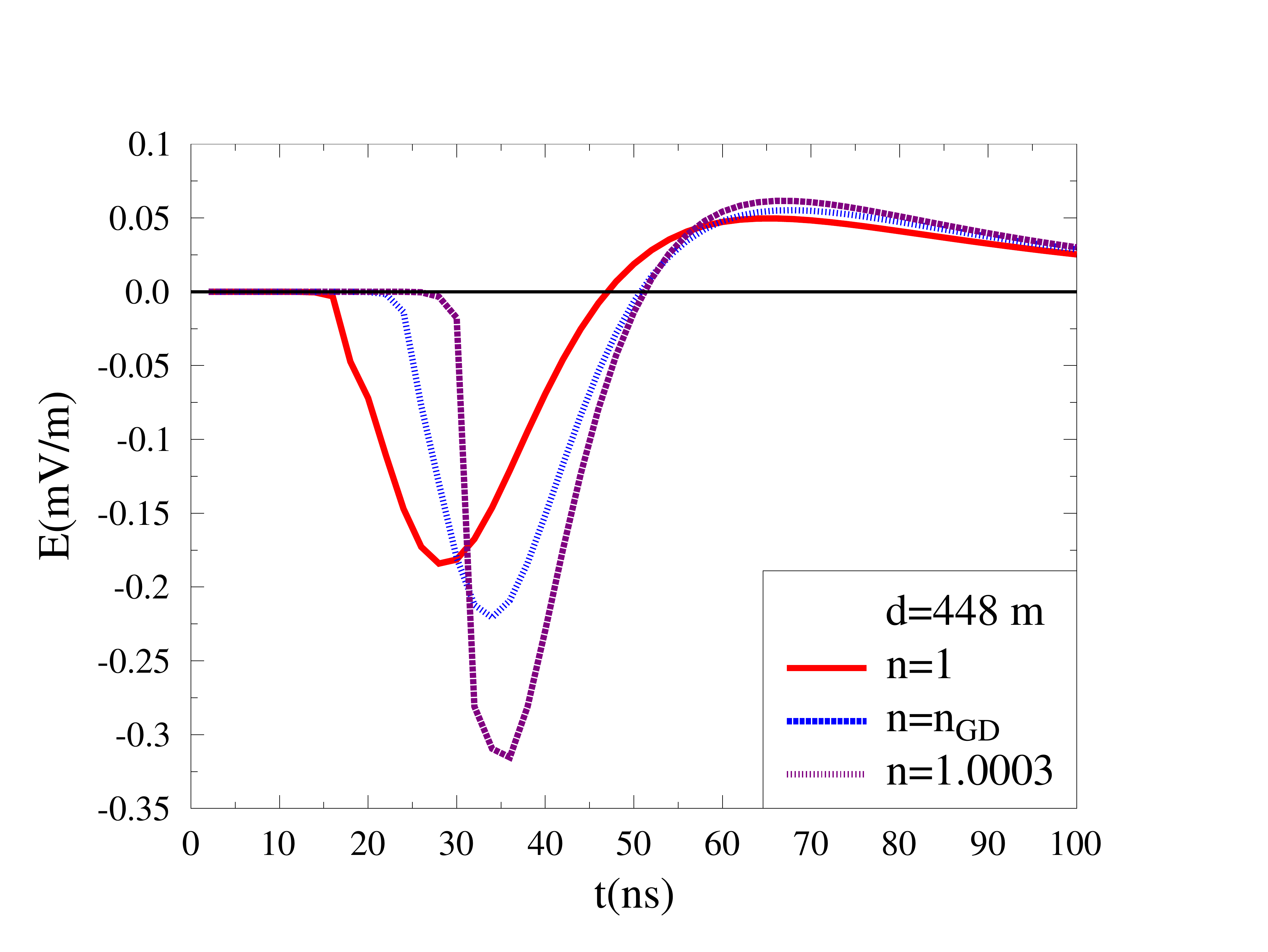}}
\caption{Same as \figref{trett} but for an observer at a distance of 448\,m from the shower core (impact distance is 400\,m). }
\figlab{trett4}
\end{figure}

The picture is rather different at a distance of 448\,m from the core (impact parameter equals 400\,m) as shown in \figref{trett4}. At this observer distance the shower velocity projected on the line-of-sight is much less than the propagation speed of radio waves, $c/n$, and as a result the influence of the air-refractivity is negligible. This is shown on the l.h.s.\ where the curves showing the mapping of shower height on observer time lie almost on top of each other. As a result also the calculated pulses for the three different cases are almost identical. The determining length scale is the longitudinal shower profile projected along the line-of-sight, compared to which the length scale of the distribution of particles in the shower core is negligible and thus can be ignored.

\begin{figure}[!htb]
\centerline{ \includegraphics[width=0.48\columnwidth, keepaspectratio]{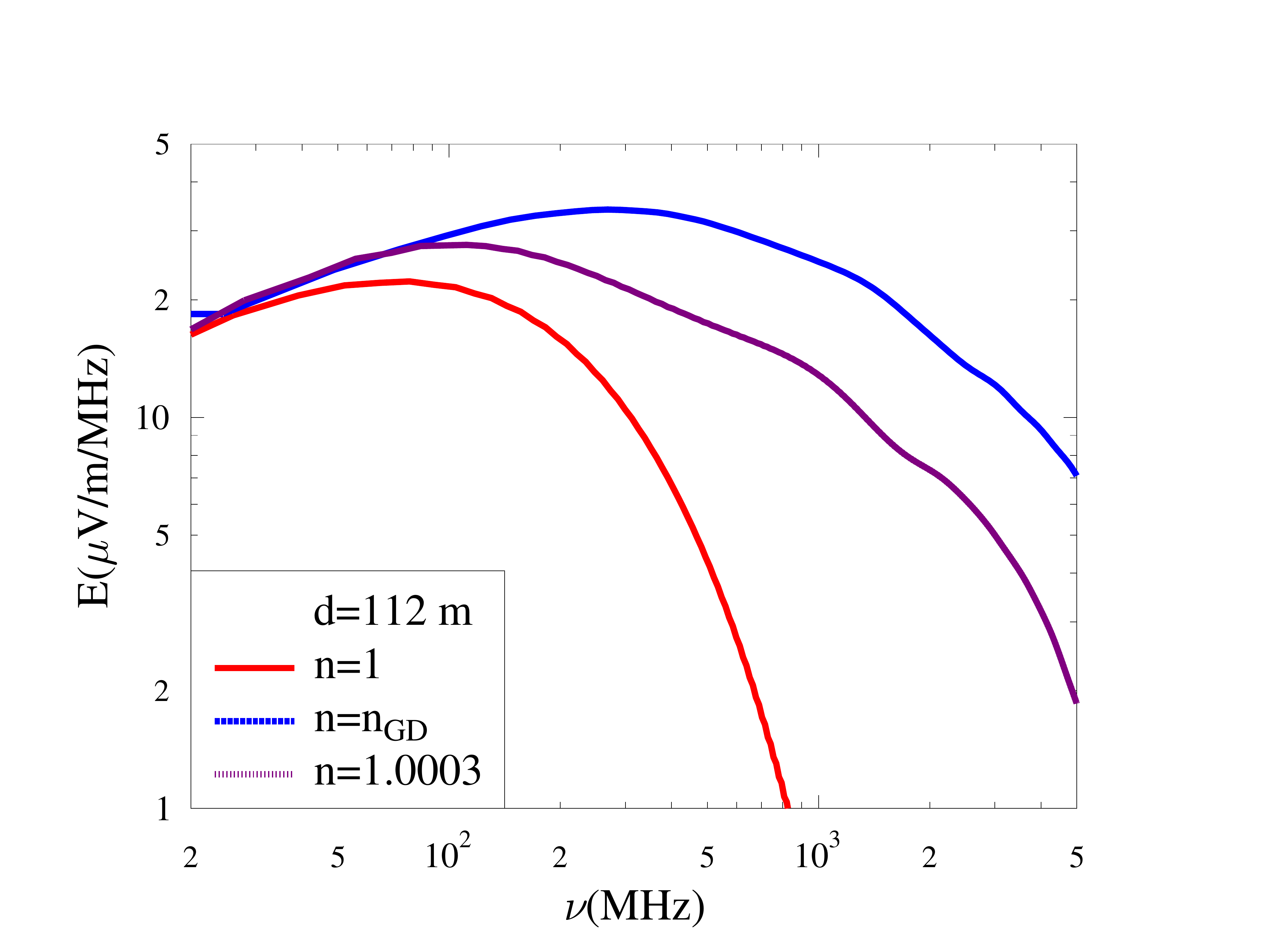}
 \includegraphics[width=0.48\columnwidth, keepaspectratio]{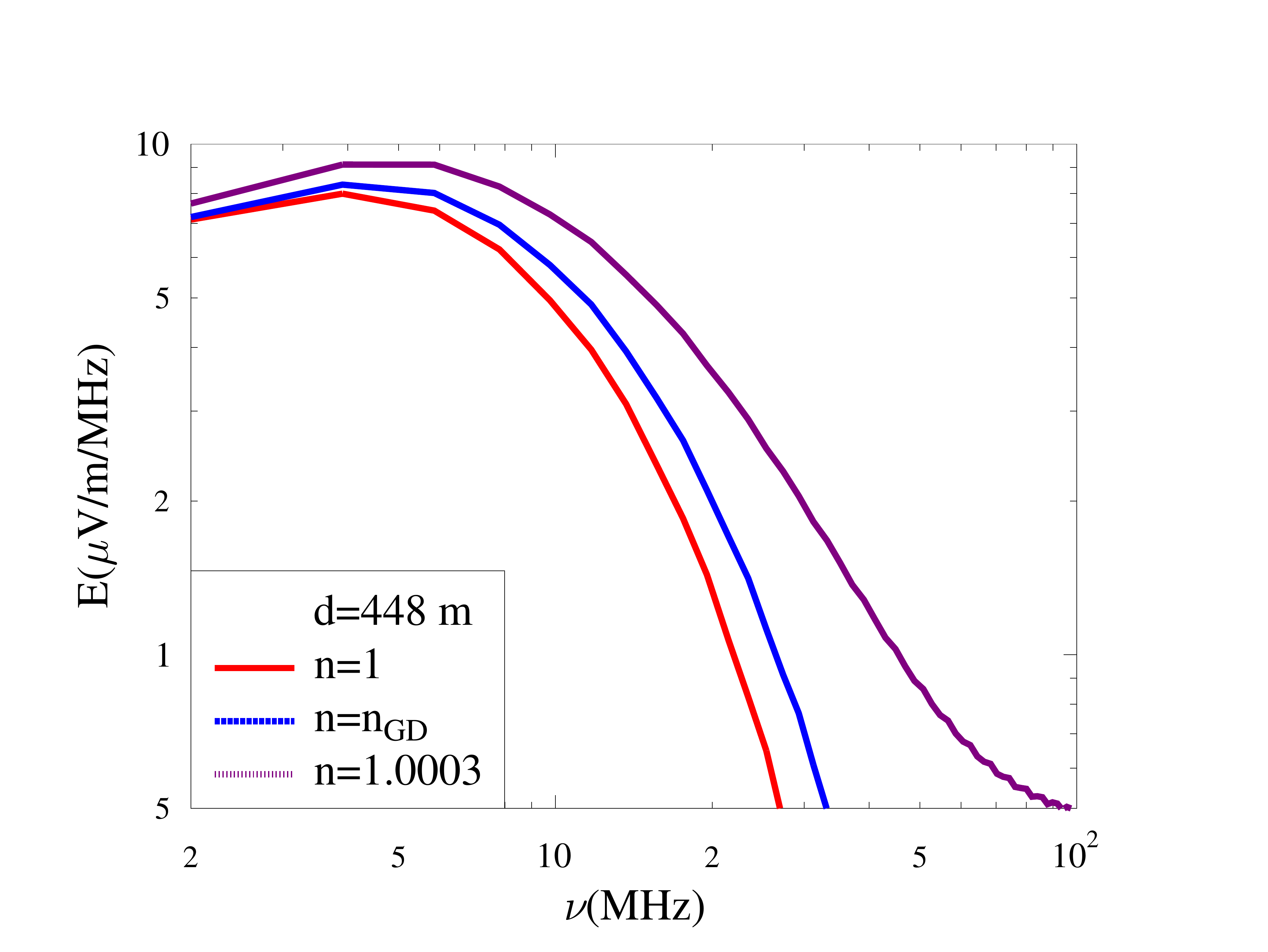}}
\caption{Frequency spectra for the calculated pulses at $d=112$\,m (l.h.s.) and $d=448$\,m (r.h.s.). }
\figlab{trettf}
\end{figure}

To obtain a better insight in the implication of the pulse width for observations, the frequency spectra are shown in \figref{trettf}. This shows that at the Cherenkov distance (defined as the distance from the shower core where the inflection point of the curve giving shower height v.s.\ observer distance coincides with the shower maximum) there are strong high frequency components in the signal ranging up to 5 GHz. This is in sharp contrast to the pulse at distances beyond the Cherenkov cone where the maximum frequency is of the order of 10's\,of MHz, dropping with distance.

The Cherenkov contribution to radio emission has distinct effects on the lateral distribution of the signal strength as is discussed in the following section. In addition we are investigating to what extent this gives us a novel and more accurate method to determine the position of the shower maximum and thus composition. An update of our previous study~\cite{dVries10} (where Cherenkov effects were not included) is forthcoming~\cite{dVries12}.

Cherenkov effects could possibly affect the separation of geomagnetic and charge excess radiation by the analysis of the polarization patters~\cite{Wer08} as discussed in~\cite{Berg12}. Even though Cherenkov effects are not changing the polarization direction of the signal, the strength or the signal at a particular distance from the shower core depends directly on the optimal emission height which may be different for the two processes. Ongoing calculations~\cite{dVries12} show that Cherenkov emission is important for certain aspect, but that the analysis of frequency filtered data, as discussed for example in~\cite{Rev11,Berg12}, is hardly affected.

\section{The lateral distribution}
\label{sec:LDF}

Recently some preliminary results were reported~\cite{Cor11} for the lateral distribution of the signal strength as measured with the antenna's in the core of LOFAR~\cite{LOFAR}, see \figref{Lofar}. LOFAR is a new generation radio-astronomy telescope which operates in the frequency regime of 30-240\,MHz. The revolutionary aspect of this telescope is that it deploys thousands of very simple wire antenna's instead of large parabolic dishes. All simple antennas can be phase-coupled to obtain the equivalent resolution of a dish with a diameter of many hundreds kilometers and a collecting area of several thousand square meters. Apart from using this telescope for astronomical observations, because of the software coupling of the individual antennas, it can also be used for observation of transient phenomena such as radio emission from air showers.

\begin{figure}[!htb]
\centerline{ \includegraphics[width=0.48\columnwidth, keepaspectratio]{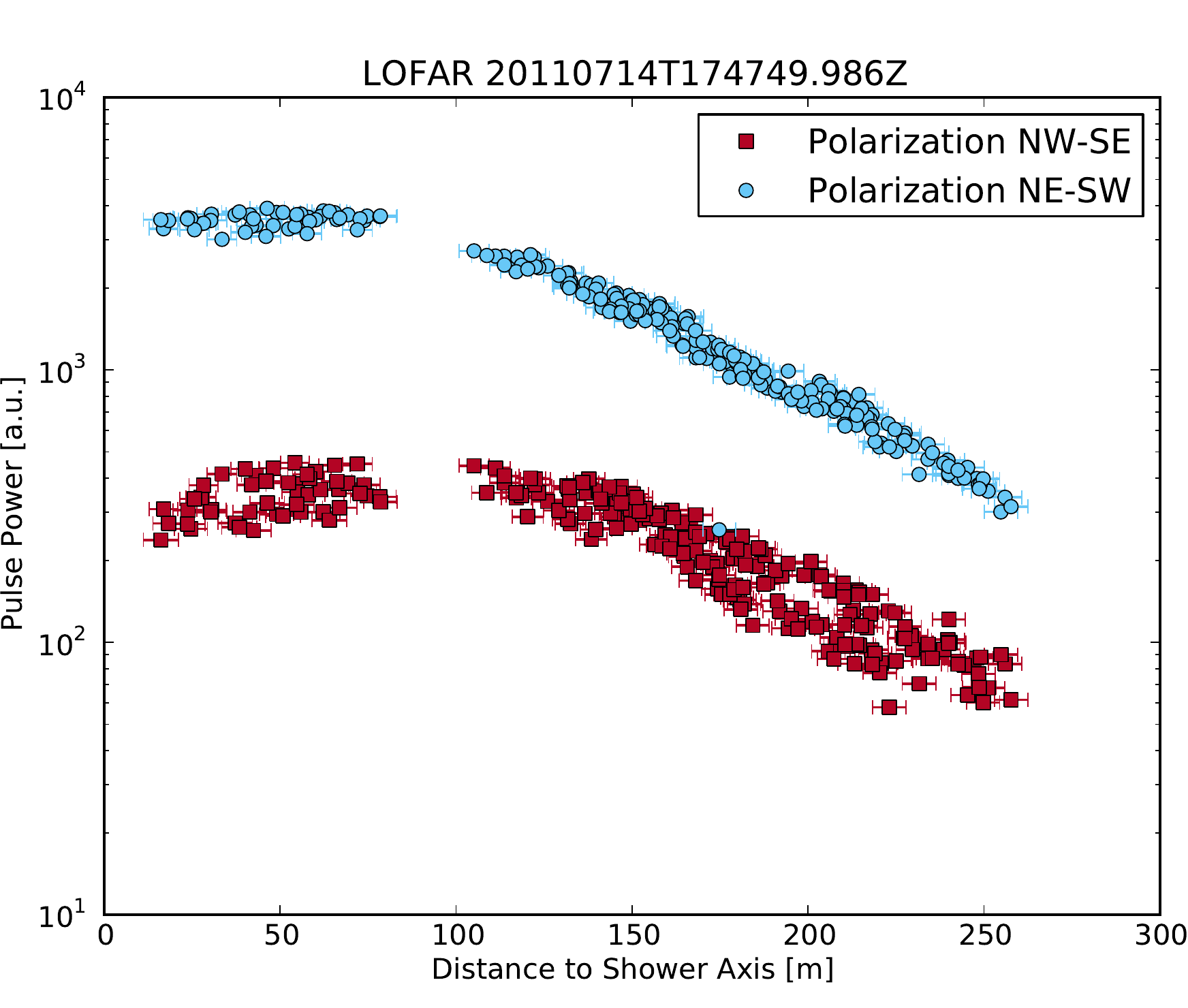} }
\caption{Figure from ref.~\cite{Cor11} showing the (un calibrated) lateral distribution function (LDF) for two polarization direction for a single event as measured with LOFAR. }
\figlab{Lofar}
\end{figure}

Due to the large number of antennas the ground pattern can be determined accurately for single showers. This is a major advantage with respect to sparse arrays where data of many showers have to be combined. The data of \figref{Lofar} have been obtained using the low-band antennas that operate in the frequency regime of 30--80\,MHz. The figure shows that for distances less than about 100\,m the intensity of the radiation is rather constant and starts to drop exponentially only at larger distances. Since the distance where the Cherenkov radiation reaches a maximum is of the order of 100\,m this structure in the LDF can be understood and it follows closely the predicted~\cite{dVries11} trend made before the experiment.

\section{Complementarity}

The strong point of cosmic-ray observations at the Pierre Auger Observatory is (besides its large surface area) the fact that it allows for hybrid observations. At present this is based on simultaneous observations with fluorescence (FD) and surface (SD) detectors where the two detectors yield complementary information on the air shower induced by the cosmic ray. Radio detection (RD) will offer additional complementarity.

Depending on the distance from the shower core, the radio pulse is sensitive to different aspects of the shower evolution. Close to the shower core, especially at the Cherenkov distance, the frequency content of the pulse is determined by the distribution of particles in the shower front~\cite{dVries11}. The Cherenkov distance itself is directly related to $X_{max}$ of the shower maximum. Therefore this yields a measurement of $X_{max}$ on an event-by-event basis, which is essential for the determination of the chemical composition. This constitutes an alternative determination of this important shower parameter, independent from FD observations. Also the fall-off of signal strength with distance to the shower core~\cite{dVries10} depends on $X_{max}$. The frequency content of the pulse at the Cherenkov angle is a direct indication of the spatial distribution of the particles in the shower core~\cite{EVA}. At present little effort has gone in the understanding of the significance of such a measurement as this could not be investigated with any other means. At larger distances the pulse shape is determined by the (derivative of) the shower profile~\cite{Sch10}. Because of the relation between shower height and observer time,  the earlier part of the shower induces the higher frequency components (which are much lower than the frequencies seen at the Cherenkov distance). This makes that the measured pulse is sensitive to the early stages of shower development which is an aspect that is not well covered by FD measurements.

In general the radio signal is exclusively sensitive to the electron content of the shower in contrast to the SD which are primarily sensitive to the muon content. In addition the polarization of the radio signal can be used to distinguish the charge excess from the geomagnetic current. No other measurement is able to do so. This will serve as an additional constraint on shower simulations.

\end{document}